\documentclass[final,12p,nopreprintline]{elsarticle}

\usepackage{subcaption}

\usepackage{graphicx}
\usepackage{framed,multirow}

\usepackage{amssymb}
\usepackage{latexsym}
\usepackage[flushleft]{threeparttable}

\usepackage{url}
\usepackage{xcolor}
\usepackage{amsmath}
\usepackage{afterpage}
\usepackage{lineno,hyperref}
\modulolinenumbers[2]

\definecolor{newcolor}{rgb}{.8,.349,.1}

\begin{document}

\begin{frontmatter}

\title{$P_2T_2$: a Physically-primed deep-neural-network approach for robust $T_{2}$ distribution estimation from quantitative $T_{2}$-weighted MRI}

\author{Hadas Ben-Atya\corref{mycorrespondingauthor}}
\ead{hds@campus.technion.ac.il}
\author{Moti Freiman}

\address{Faculty of Biomedical Engineering, Technion - Israel Institute of Technology, Haifa Israel}
\cortext[mycorrespondingauthor]{Corresponding author}
\begin{abstract}
Estimating $T_2$ relaxation time distributions from multi-echo $T_2$-weighted MRI ($T_2W$) data can provide valuable biomarkers for assessing inflammation, demyelination, edema, and cartilage composition in various pathologies, including neurodegenerative disorders, osteoarthritis, and tumors. Deep neural network (DNN) based methods have been proposed to address the complex inverse problem of estimating $T_2$ distributions from MRI data, but they are not yet robust enough for clinical data with low Signal-to-Noise ratio (SNR) and are highly sensitive to distribution shifts such as variations in echo-times (TE) used during acquisition. Consequently, their application is hindered in clinical practice and large-scale multi-institutional trials with heterogeneous acquisition protocols. We propose a physically-primed DNN approach, called $P_2T_2$, that incorporates the signal decay forward model in addition to the MRI signal into the DNN architecture to improve the accuracy and robustness of $T_2$ distribution estimation. We evaluated our $P_2T_2$ model in comparison to both DNN-based methods and classical methods for $T_2$ distribution estimation using 1D and 2D numerical simulations along with clinical data.  Our model improved the baseline model's accuracy for low SNR levels ($SNR<80$) which are common in the clinical setting. Further, our model achieved a $\sim$35\% improvement in robustness against distribution shifts in the acquisition process compared to previously proposed DNN models. Finally, Our $P_2T_2$ model produces the most detailed Myelin-Water fraction maps compared to baseline approaches when applied to real human MRI data. Our $P_2T_2$ model offers a reliable and precise means of estimating $T_2$ distributions from MRI data and shows promise for use in large-scale multi-institutional trials with heterogeneous acquisition protocols.
Our source code is available at: \\\url{https://github.com/Hben-atya/P2T2-Robust-T2-estimation.git}.

\end{abstract}

\begin{keyword}
\texttt Deep-Learning\sep $T_2$ Relaxometry\sep Myelin-Water Imaging\sep Microstructure Imaging \sep Physically-driven deep-learning \\
\vfill{}
\end{keyword}

\end{frontmatter}





\section{Introduction}
Estimating $T_2$ relaxation distribution in tissue from $T_2$-Weighted MRI data acquired at multiple echo times (TEs) presents a promising approach to independently characterize different water pools within the tissue. While free water displays pure monoexponential decay with long $T_2$ values, water in tissue bound to lipids and proteins exhibits distinct relaxation behavior with shorter $T_2$ values \citep{Fatemi2020FastApproach, MargaretCheng2012PracticalRelaxometry}. In the context of neuroimaging, such water pools can include the Myelin sheath, intra- and extra-axonal space (IES), the gray matter, the cerebrospinal fluid (CSF) and pathological tissues \citep{MIML}. 

$T_2$ distribution mapping has a wide range of clinical applications, including aiding in the identification of areas of inflammation and demyelination in the brain and spinal cord \citep{Lee2020QuantitativeModel, Vymazal1999T1Content}. It can also be used to aid in the diagnosis and treatment planning of neurodegenerative diseases such as multiple sclerosis \citep{Mueller2007Voxel-basedSclerosis}, stroke \citep{Bauer2014QuantitativeStroke, Duchaussoy2019SyntheticManagement}, and epilpngy \citep{Winston2017AutomatedEpilpngy}. Furthermore, $T_2$ distribution mapping can be used to assess musculoskeletal disorders by characterizing cartilage degeneration in osteoarthritis \citep{Verschueren2021T2Reproducibility}. In oncology, this technique can assist in tumor characterization \citep{Cieszanowski2012CharacterizationTimes, Dregely2016Rapid3T, Yamauchi2015ProstateSequence}.

Specifically, Bontempi et al. \citep{Bontempi2021QuantitativeGliomas} demonstrated the added value of $T_2$ distribution mapping in detecting subtle alterations at the periphery of lower grade Gliomas by decomposing the $T_2$ distribution into three main components: myelin-associated water, intra/extracellular water, and cerebrospinal fluid. This technique proved more effective than conventional scalar $T_2$-mapping. Raj et al. \citep{Raj2014Multi-compartmentModel} and Chatterjee et al. \citep{chatterjee2018identification} have also used $T_2$ distribution mapping to precisely characterize the demyelination process of white-matter lesions in multiple sclerosis patients. In oncology, Nikiforaki et al. \citep{ Nikiforaki2020MultiexponentialTumours} employed $T_2$ distribution mapping to discriminate between benign and malignant adipocytic tumors.

Yet, accurate estimation of the $T_2$ distribution can be challenging due to the requirement of a high signal-to-noise ratio (SNR), which may not be feasible in clinical settings due to long acquisition times \citep{Canales-Rodriguez2021RevisitingSquares}. Additionally, classical methods for $T_2$ distribution mapping can incur a heavy computational burden due to the application of either regularized non-linear least-squares (NNLS) optimization \citep{NNLS1, decaes} or a dot-product search to estimate the $T_2$ distribution \citep{Radunsky2021QuantitativeTime}.

A supervised deep neural network (DNN) approach was recently proposed for accelerating the estimation of the $T_2$ distribution \citep{MIML, SynthMap, Liu2020MyelinMinuteb}. However, the requirement of high SNR data makes it difficult to apply these DNN-based approaches in clinical settings. Moreover, these methods are vulnerable to various distribution shifts, including changes in MRI acquisition protocols such as echo-time (TE), limiting their ability to generalize compared to classical NLLS approaches, which are insensitive to specific TE settings. The susceptibility to distribution shifts in acquisition protocols also practically precludes the use of DNN-based approaches in large-scale clinical trials with diverse MRI acquisition protocols \citep{Verschueren2021T2Reproducibility}.

The main goal of this study is to enhance the ability of DNN-based methods for estimating $T_2$ distributions from $T_2$-weighted MRI data acquired with low SNR and with different echo times (TEs), and to make them more resistant to distribution shifts. We propose to achieve this goal by explicitly incorporating the TEs used during MRI acquisition into the DNN architecture and using this information during inference. This will enable the DNN to better generalize the ill-posed inverse problem that is associated with $T_2$ mapping from MRI data. Thus, making it more applicable in the clinical setting and in large-scale clinical trials with diverse MRI acquisition protocols.

\begin{figure*}[th!]
\centering
\includegraphics[width=0.99\textwidth]{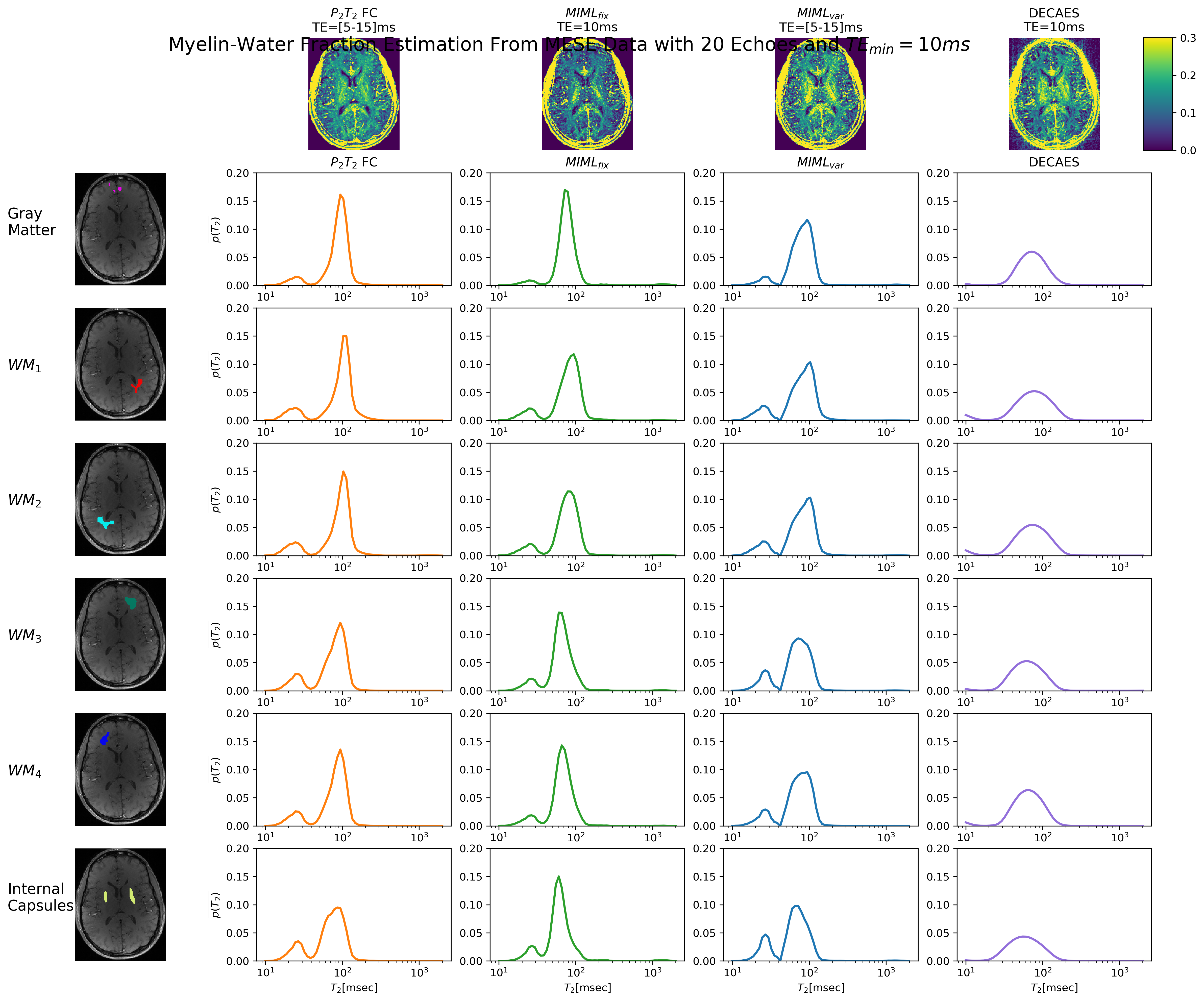}
\caption{
Estimation of Myelin-Water Fraction from Multi-Echo Spin Echo MRI Scans using the Blue Dataset \citep{Whitaker2020MyelinMRI}. The top row of figures shows the predicted Myelin-Water Fraction (MWF) maps for each of the four models. The subsequent rows (2-7) depict the first echo from the Multi-Echo Spin Echo (MESE) scan with the corresponding region of interest (ROI), followed by the $p(T_2)$ predictions for each voxel within the ROI and their average prediction for the four models. Among the models tested, the $P_2T_2$-FC model showed superior performance by producing a smooth MWF map and detailed $T_2$ distributions, where the myelin component was distinguishable from the other components.
}
\label{fig:blue_data}
\end{figure*}

The primary outcome of our study is illustrated in Figure~\ref{fig:blue_data}. The figure shows a comparison of our $P_2T_2$ model's myelin-water fraction (MWF) map prediction for an MRI scan of a healthy subject \citep{Whitaker2020MyelinMRI} against a baseline DNN-based MIML model \citep{MIML} trained using: 1) varying acquisition sequences ($TE \in [5,15]$ms), and 2) a fixed acquisition sequence ($TE=10$ ms), and the classical DECAES \citep{decaes} methods. Our $P_2T_2$ model produced a less noisy and more detailed MWF map in comparison to the other methods.

Our study makes several significant contributions to the field of $T_2$ distribution estimation:
\begin{itemize}
    \item Introducing a physically-informed approach that incorporates the TEs used during MRI data acquisition
    \item Demonstrating that incorporating the TEs into the DNN architecture enhances the generalization capacity and robustness of the DNN-based method for $T_2$ distribution estimation.
    \item Proposing a method for generating realistic multi-echoes MRI data based on real brain segmentation.
    \item  Conducting extensive evaluations of our physically-informed approach compared to existing methods on both synthetic and real MRI data from healthy patients.
    \item  Publicly available  source code package for $T_2$ distribution estimation: \url{https://github.com/Hben-atya/P2T2-Robust-T2-estimation.git}
\end{itemize}


The rest of this paper is structured as follows: in Section \ref{sec:back}, we provide an overview of $T_2$ distribution mapping from $T_2W$ MRI data. Section \ref{sec:methods} describes our proposed $P_2T_2$ approach, which incorporates the TEs used during MRI data acquisition into the DNN architecture. In Section \ref{sec:eval}, we detail our evaluation methodology. Section \ref{sec:results} presents the results of our evaluation, including comparisons between our proposed approach and existing methods on both synthetic and real healthy-patient MRI data. Section \ref{sec:discussion} discusses the experimental results, indicates some limitations of our study, and concludes our work while highlighting future directions for research in this area.

\section{Background}
\label{sec:back}

The $T_2$ distribution is typically estimated by fitting a signal-decay model to MRI data obtained at multiple echo times ($TE$). This fitting process can be formulated as a regularized least-squares problem, which can be written as:
\begin{equation}
\widehat{p(T_2)} = \arg\min_{p(T_2)} \sum_i^{N}||s(TE_i) - s_i||_2 ^2 + \lambda \mathcal R,
\label{eq:nnls_optim_func}
\end{equation}
where $\widehat{p(T_2)}$ is the estimated $T_2$ distribution, $s_i$ is the measured signal at $TE_i$, and $s(TE_i)$ is the expected signal at $TE_i$. The regularization term $\mathcal R$ is included to mitigate overfitting of the model to the data and ensure that the estimated $T_2$ distribution is smooth and physically plausible.

To compute the expected signal $s(TE_i)$ from the estimated $T_2$ distribution, we use the extended phase graph (EPG) mechanism \citep{EPG}. This mechanism takes into account several variables, including the echo time ($TE_i$), the relaxation times $T_1$ and $T_2$, the flipping angle $\alpha$, and the probability distribution of $T_2$, denoted as $p(T_2)$:

\begin{equation}
s(TE_i) = \int EPG\left( TE_i, T_1, T_2, \alpha \right)p(T_2) , dT_2
\label{eq:the_MRI_signal}
\end{equation}

Here, $EPG(TE_i, T_1, T_2, \alpha)$ is a function that models the behavior of the MRI signal given the input parameters. The integral in Eq.~\ref{eq:the_MRI_signal} represents the expected signal obtained by integrating the EPG mechanism over the range of $T_2$ values.

Unfortunately, Eq.~\ref{eq:nnls_optim_func} cannot be optimized directly, making it impractical. To simplify the problem, parametric approaches are used to represent the $T_2$ distribution $p(T_2)$ as a linear combination of a fixed number of water pools:

\begin{equation}
p(T_2) = \sum_{i=1}^n v_iF_i(m_i,T_2)
\label{eq:parametric}
\end{equation}

This representation uses $n$ water-pools, where $v_i$ represents the volume fraction of the $i^\textrm{th}$ component, $F_i$ is the probability distribution function (pdf), and $m_i$ are the distribution parameters.

The number and types of water-pools used in these models are often based on prior knowledge, such as myelin water, the water in the intra/extracellular space (IES), and cerebrospinal fluid (CSF). These water-pool probability distribution functions can be represented by various parametric distributions, including Delta and Gaussian distributions, as detailed in \citep{Bai2014AImages,Raj2014Multi-compartmentModel}.

However, assumptions made about the number of water pools, their associated $T_2$ ranges, and pdfs may not accurately reflect the tissue microscopic content, especially in pathological tissue. As a result, such assumptions can lead to suboptimal results. Additionally, accurately calculating the expected signal using Eq.~\ref{eq:the_MRI_signal} can be computationally burdensome, making practical utilization challenging \citep{Canales-Rodriguez2021RevisitingSquares}.

Non-parametric approaches aim to overcome the issues associated with parametric approaches by avoiding prior assumptions about the $T_2$ distribution. These approaches discretize Eq.~\ref{eq:the_MRI_signal} as a product of a dictionary matrix and a discretized distribution. The $T_2$ distribution is then estimated by solving a regularized non-negative least squares (NNLS) problem \citep{NNLS1}.


Dictionary-based methods utilize a dictionary search to bypass the optimization process. In these methods, a dictionary of expected signals is created using Bloch simulations, based on an assumed range of water-pools, their fractions, and parameters, as well as the $T_2$ distribution used to generate the signal. The best-matching signal in the dictionary is determined by minimizing the distance function between the observed and expected signal. Subsequently, the matched signal's $T_2$ distribution is utilized \citep{Radunsky2021QuantitativeTime}.



However, these methods have limitations in clinical settings due to a requirement for a high signal-to-noise ratio (SNR), which can be difficult to achieve within practical acquisition times. Additionally, both approaches can be computationally demanding, with the need to apply either regularized NNLS optimization \citep{NNLS1,decaes} or a dot-product search for $T_2$ distribution estimation \citep{Radunsky2021QuantitativeTime}.


Recently, to overcome the limitations of classical methods, several researchers have proposed a supervised deep neural network (DNN) approach for accelerating the estimation of the $T_2$ distribution \citep{MIML, SynthMap, Liu2020MyelinMinuteb}. In this approach, the $T_2$ distribution estimation is defined as a function of the DNN weights $\theta$ and the forward pass of the DNN $\mathcal F_\theta$ applied to the observed signal $s$ at multiple TEs, with the output $y$ being the estimated $T_2$ distribution.
\begin{equation}
\hat{y} = F_\theta \left(s \right)
\label{eq:miml_formula}
\end{equation}
To train the DNN model, simulated data are used, where a pre-defined range of components and their fractions are considered, and reference $T_2$ distributions are generated using a random generator, along with the corresponding MRI signal for each TE in the multi-echo acquisition using Eq.~\ref{eq:the_MRI_signal}. The DNN weights are then obtained using a supervised learning scheme, where the loss function penalizes differences between the DNN estimation and the target.
\begin{equation}
\hat{\theta} = \sum_{i=1}^N L \left(y_{ref},F_\theta \left(s \right)\right)
\label{eq:miml_train_formula}
\end{equation}

Model-informed machine learning (MIML) \citep{MIML} is one such method that generates a non-parametric representation of the $T_2$ distribution using a DNN. However, MIML relies on a semi-parametric approach to simulate reference $T_2$ distributions, which sets the number of water-pools and associated parameters based on prior assumptions. Another DNN approach, SynthMap \citep{SynthMap}, predicts water-pool component fractions, such as MWF maps, using a CNN model trained on simulated realistic MRI data based on brain templates. This approach improves prediction accuracy and precision by leveraging spatial correlations between the pixels. However, it does not provide a detailed characterization of the entire $T_2$ distribution.


 Yet, the susceptibility of DNN-based methods to distribution shifts in acquisition protocols and the requirement for high SNR signals  practically precludes the use of DNN-based approaches in large-scale clinical trials with diverse MRI acquisition protocols.


\section{Methods}
\label{sec:methods}
\subsection{Deep-neural-network model}
\begin{figure*}[t!]
    \centering
    \includegraphics[width=\textwidth]{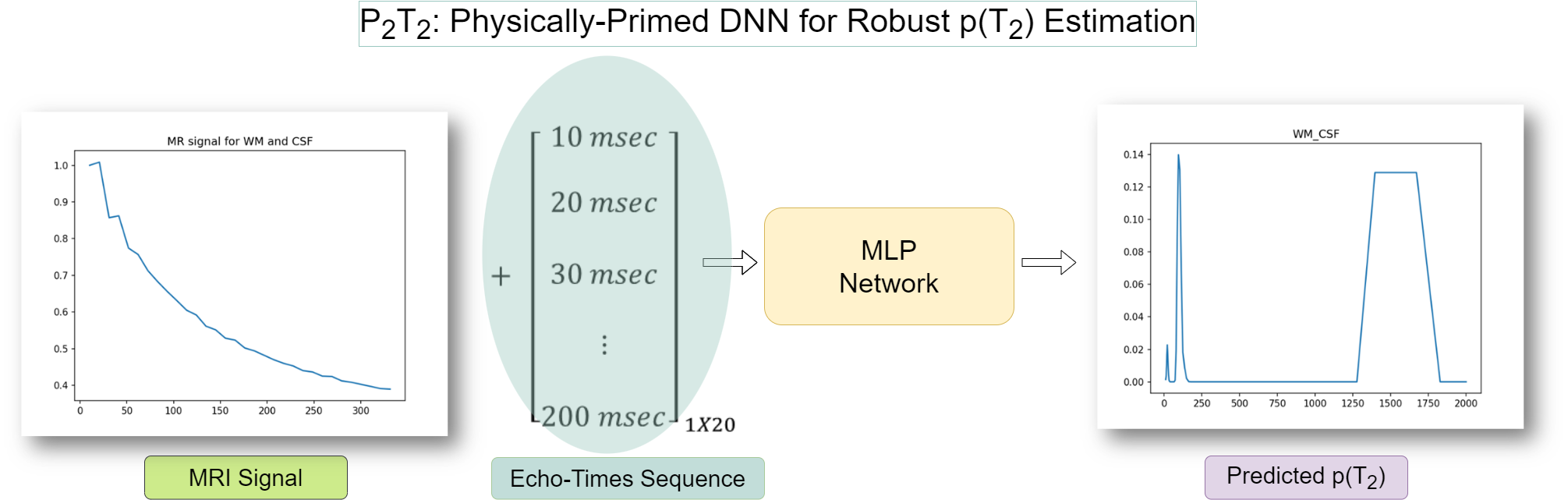}
    \caption{
    The $P_2T_2$ model incorporates echo-time sequence information as additional input, which is encoded into the model architecture (highlighted in green). The inputs to the model include both the MRI signals acquired at different TEs and the corresponding TEs used during acquisition. The output of the model is an estimation of the discretized $T_2$ distribution.  
    }
    \label{fig:$P_2T_2$_added_value}
\end{figure*}

We propose a physically-informed deep neural network (DNN) model, called $P_2T_2$, that incorporates echo-time (TE) acquisition parameters into the model architecture alongside magnetic resonance imaging (MRI) signals.  
Figure \ref{fig:$P_2T_2$_added_value} presents our main contribution compared to the previously proposed MIML model \citep{MIML}. The additionally encoded echo-times are highlighted with green. 

The $T_2$ distribution estimation problem is formally defined as follows:

\begin{equation}
\hat{p}(T_2) = F_\theta \left(s, TEs \right)
\end{equation}

In this equation, $\hat{p}(T_2)$ represents the discretized predicted $T_2$ distribution, while $F_\theta$ is our DNN model with learnable weights $\theta$. The variable $s$ represents the $T_2W$ signals, and $TEs$ are the echo-times utilized during data acquisition. Our primary architecture consists of a linear fully-connected multi-layer model.

We have devised two distinct approaches for incorporating the TEs (echo times) into our deep neural network (DNN) model.

Our principal model, the $P_2T_2$-FC model, consists of twelve linear layers, each with 256 neurons, along with an output layer comprising 60 neurons that map the predicted signal to the discretized $T_2$ distribution size. The $P_2T_2$-FC model accepts a 1D signal as input, where the TE sequence is concatenated with the observed MRI signal. Hence, the input layer for our model with a 20-echoes MRI signal consists of a linear layer with 40 neurons. All layers, except for the output layer, utilize the Rectified Linear Unit (ReLU) activation function, while the output layer follows a SoftMax activation function to convert the predicted signal into a distribution signal.

Our second alternative, the $P_2T_2$-Convolution-FC, operates on the MRI signals and TE arrays as a 2D input array. A 2D convolution layer is employed as an input layer to transform the 2D input signal into a 1D signal. This is followed by six linear layers, each consisting of 256 neurons, and an output layer of 60 neurons. Similar to the $P_2T_2$-FC model, the ReLU function is employed as the activation function for each layer, while the output layer follows a SoftMax layer.

The $P_2T_2$-FC model boasts of 744,448 trainable parameters, whereas the $P_2T_2$-Convolution-FC model only requires 349,699 parameters. However, both models, as well as the baseline model, exhibit nearly identical inference times, as demonstrated in Table ~\ref{table:models_parameters}.

The training procedure for both models is as follows:

\begin{equation}
\hat{\theta} = \sum_{i=1}^n L \left(y_{ref},F_\theta \left(s, TEs\right)\right)
\end{equation}

In this equation, $\hat{\theta}$ represents the learned weights of the DNN model, while $y_{ref}$ is the reference $T_2$ distribution. The loss function $L$ measures the difference between the predicted $T_2$ distribution and the reference distribution. We optimize the DNN model by minimizing this loss function with respect to the weights.

We trained our models on a synthetic dataset simulated with various TEs to promote the generalization of our DNN model in tackling the ill-posed problem of $T_2$ distribution estimation. This approach discourages the DNN model from solely learning the correlation between the input and output data and instead encourages it to learn the underlying physics of the problem \citep{Avidan2022Physically-primedReconstruction}.



We used a combination of the mean-squared error (MSE) and the Wasserstein distance as the loss function \citep{MIML}: 

\begin{equation}
\mathcal{L} = L_{MSE} + \lambda * L_{Wasserstein}
\end{equation}

In this equation, $L_{MSE}$ represents the mean-squared error loss function, and $L_{Wasserstein}$ represents the Wasserstein distance between the predicted and ground truth $T_2$ distributions. The Wasserstein distance is a distance function that measures the movement required to transform one probability distribution into another, penalizing any significant deviation between the predicted and ground truth $T_2$ distributions. The MSE component of the loss function ensures that the predicted $T_2$ distribution is similar to the ground truth distribution. The hyperparameter $\lambda$ controls the weight of the Wasserstein distance term in the overall loss function. By incorporating both MSE and Wasserstein distance terms, we ensure that our DNN model generates $T_2$ distributions that are not only accurate but also consistent with the ground truth.

\subsection{Synthetic Data generation}
\label{sec:Data_generation}



We generated synthetic data following the methodology proposed by Yu et al. \citep{MIML} to train our models. Specifically, we modeled the $T_2$ distribution as a mixture of Gaussian distributions, where each Gaussian corresponds to a different water pool component. The probability density function of the mixture model is defined as:

\begin{equation}
p(T_2) = \sum_{i=1}^{n}{\frac{v_i}{\sigma_i \sqrt{2\pi}}\exp\left(-\frac{(T_2 -\mu_i)^2}{2\sigma_i^2}\right)}
\label{eq.gaussian_dist}
\end{equation}

In this equation, $v_i$ denotes the volume fraction of the $i$th component, which is randomly sampled from a Dirichlet distribution subject to the constraint that $\sum_{i=1}^{n}{v_i}=1$. The mean ($\mu_i$) and standard deviation ($\sigma_i$) of each component are randomly drawn from a uniform distribution with bounds specified in Table~\ref{table:components_$T_2$}. We use $n$ to represent the number of components in the mixture model. By simulating synthetic data in this manner, we can generate diverse and complex $T_2$ distributions that cover a wide range of possible scenarios, allowing our DNN model to learn the underlying physics of the problem and generalize to real-world scenarios.


We categorized $T_2$ distributions in the brain into seven different combinations, each consisting of a mixture of up to three water pools \citep{MIML}:

\vspace{-\topsep}

\begin{itemize}
\setlength{\parskip}{0pt}
\setlength{\itemsep}{0pt plus 1pt}
\item White matter (WM): Myelin and IES water pools.
\item Gray matter (GM): Myelin and GM water pools.
\item Cerebrospinal fluid (CSF): CSF water pools.
\item Mixture of WM and GM: Myelin, IES, and GM water pools.
\item Mixture of WM and CSF: Myelin, IES, and CSF water pools.
\item Mixture of GM and CSF: Myelin, GM, and CSF water pools.
\item Pathology: Pathology water-pool.
\end{itemize}

Table~\ref{table:components_$T_2$} provides a summary of the parameters of the different components \citep{MIML}.

\begin{table}[h!]
\caption{\label{table:components_$T_2$}The water-pools characterization used for the Gaussian,$T_2$ distributions for the data simulation}
\centering
\begin{tabular}{|p{2.25cm}|p{2cm}|l|p{4cm}|p{3cm}|p{2cm}|}
\hline
Water Pool  & Range of $T_2$ Mean($\mu$) &  Range of $T_2$ Std.($\sigma$) \\
\hline
Myelin    & 15-30 ms     & 0.1-5 ms \\
IES       & 50-120 ms    & 0.1-12 ms \\
GM        & 60-300 ms    & 0.1-12 ms \\
Pathology & 300-1000 ms  & 0.1-5 ms \\
CSF       & 1000-2000 ms & 0.1-5 ms \\
\hline
\end{tabular}
\end{table}

We discretized the $T_2$ distribution using a densely sampled grid with 1ms resolution to enable numerical computations. We then simulated the $T_2$-weighted MRI signals by multiplying the EPG matrix with the discretized $T_2$ distribution, and summing over $T_2$ values using the following equation:

\begin{equation}
s(TE_i) = \sum_{T_2} EPG\left( TE_i, T_1, T_2, \alpha \right) * p(T_2)
\label{eq.mri_signal}
\end{equation}

We randomly sampled TEs from a range of [5ms, 15ms]  and varied $\alpha$ from $90^{\circ}$ to $180^{\circ}$ with uniform distributions to enhance the robustness of our models against variations in TE spacings and flip angles. We generated a total of 1,400,000 sets of $T_2W$ signals corresponding to the 7 water pool combinations using the $T_2$ distribution.

We downsampled the densely discretized $T_2$ distribution, consistent with the approach employed by classic methods and previous DNN models \citep{MIML} to generate reference distribution data.

\subsection{Implementation details}

We implemented our models and loss functions using PyTorch 1.8 and Python 3.8. The training and evaluation of the models were conducted on an NVIDIA Tesla driver with CUDA 11.3. We utilized an Adam optimizer with a learning rate of $1.0e-4$ to train each model for a total of 1000 epochs. We chose the epoch with the lowest validation loss as the final model.

\begin{table}[h!]
\caption{\label{table:models_parameters}Architecture Parameters}
\centering
\begin{tabular}{|l|l|l|l|}
\hline
 data & $P_2T_2$ FC & $P_2T_2$ Convolution-FC & MIML \\
 \hline
Num. parameters &  744,448  & 349,699  & 349,696 \\
\hline
Num. layers &  12 FC & 6 FC and 1 convolution &  6 FC \\
\hline
Average time*  & 0.93 sec    & 0.8 sec  & 0.7 sec \\
\hline
\end{tabular}

\begin{tablenotes}
  \small
  \item * Average time for computing $T_2$ distribution image from a 2D multi-echo spin-echo sequence with the following parameters: acquisition matrix = 176x208; voxel-size = 1.x1.; number of echoes (N-echoes) = 20; using an NVIDIA Tesla driver with CUDA 11.3.
\end{tablenotes}

\end{table}

\section{Evaluation}
\label{sec:eval}
\subsection{Compared models}
We assessed the effectiveness of our physically-primed $P_2T_2$-FC and $P_2T_2$-Convolutional-FC models by comparing them to classical approaches, such as the DECAES algorithm \citep{decaes} and the NLLS algorithm with Laplacian regularization, as well as to an in-house implementation of the MLP-based MIML model proposed by \citep{MIML}. The MIML model's implementation was validated using the data provided in the original publication \citep{MIML}. The different DNN models were initially trained using synthetic data generated according to the method described in Section~\ref{sec:Data_generation}, with $SNR$ values ranging from 80 to 200 and $TE_{min} = \Delta TE$ values ranging from 5 to 15 ms. The following architectures were employed during training:

\begin{itemize}
\item Our $P_2T_2$-FC model, which featured 12 fully-connected layers.
\item Our $P_2T_2$-Convolutional-FC model, which consisted of a convolutional layer followed by six fully-connected layers.
\item The MIML model, which comprised of six fully-connected layers (MIML$_\textrm{var}$).
\end{itemize}

Furthermore, we also trained the MIML$_\textrm{fix}$ model, which consisted of six fully-connected layers and used simulated signals with $SNR$ values ranging from 80 to 200 and a fixed $TE{min} = \Delta TE$ value of 12 ms, as described in \citep{MIML}.

\subsection{Models accuracy in unseen SNR levels}
\subsubsection{One-dimensional simulated data}
\label{sec:fix_te_1d_test}


We first evaluated the accuracy of $T_2$ distribution produced by the $P_2T_2$ models trained under various echo-time configurations, in comparison to the prediction accuracy of the MIML$_\textrm{fix}$ model trained solely on a single acquisition protocol. 
We generated a test set consisting of 140,000 $T_2W$ MRI signals, with 20,000 signals for each of the seven TE combinations, using the methodology outlined in Section~\ref{sec:Data_generation}.
We used an echo-train with $TE_{min}=\Delta TE = 12ms$, same as those used to train the MIML$_\textrm{fix}$.
In order to evaluate the models' accuracy under low-unseen SNR levels, we created several test sets by adding Rician noise to each set at relative levels of [10, 20, 30, 40, 80, 150, 200, 400, 1000] with respect to the first echo.

We utilized the various DNN models to predict the $T_2$ distribution $p(T_2)$ from the signals. The predicted $p(T_2)$ was then compared to the reference distributions using the mean-squared error (MSE) and Wasserstein distance.

\subsubsection{Two-dimensional brain-based data}

We further evaluated the accuracy of the different models on a realistic brain-based dataset, using brain segmentation from OASIS dataset \citep{OASIS}, including the White Matter (WM), Grey Matter (GM), and CSF. 

We applied Gaussian smoothing ($\sigma=1$) to each tissue (WM, GM, and CSF), which resulted in a continuous transition between the different brain tissues to simulate the partial volume effect. We then used the smoothed tissue value as the tissue's volume fraction ($v_t$).

We calculated the water-pools' fractions ($v_i$) for each voxel and label as follows. We first divided each label map into 8x8 pixel patches. We assigned volume fractions for each patch by randomly sampling from a uniform distribution within the appropriate ranges (Table~\ref{table:water_pools_vf}). Finally, we applied a Gaussian filter ($\sigma=3$) to the volume fraction maps of each label.

\begin{table}[h!]
\caption{\label{table:water_pools_vf}The water-pools' volume fraction used for the brain-data simulation}
\centering
\begin{tabular}{|l|l|l|}
  \hline 
  Brain Tissue & Water Pool  & Range ($v_i$)  \\
  \hline 
  \multirow[c]{3}{*}[0in]{WM} & Myelin ($v_M$) & [0, 0.4] \\
  & IS ($v_{IS}$) & [0, 0.6] \\ 
  & ES ($v_{ES}$) & $1 - v_M - v_{IS}$ \\
  \hline
  \multirow[c]{3}{*}[0in]{GM} & Myelin ($v_{M}$) & [0, 0.05] \\
  & GM ($v_{GM}$)& $1-v_{M}$ \\
\hline
CSF & CSF ($v_{CSF}$) & 1 \\
\hline
\end{tabular}
\end{table}

We calculated the final $p(T_2)$ distribution maps as a weighted sum of the tissues' fractions ($v_t$) and the water-pools' fractions ($v_i$):

\begin{equation}
p(T_2) = \sum_{t}{v_t\sum_{i}{\frac{v_i}{\sigma \sqrt{2\pi}}exp\left( \frac{-(T_2 -\mu_i)^2}{2\sigma_i^2}\right)}}
\label{eq.oasis_gaussian_dist}
\end{equation}

\begin{equation*}
v_t, v_i \in \left[0,1\right], \sum_{i}{v_t}= \sum_{i}{v_i}=1
\label{eq.vf}
\end{equation*}

We generated the MRI EPG using a fixed TE array of 12ms to 240ms and a random flipping angle ($\alpha$). We then simulated multi-echo $T_2W$ signals by multiplying the EPG matrix with the computed $p(T_2)$, as described in sec.~\ref{sec:Data_generation}. We examined the models' accuracy under varying levels of noise, by adding Rician noise. We predicted $p(T_2)$ from the resulting signals using both MIML$_{fix}$ and $P_2T_2$-FC models.

We evaluated the accuracy of the different models' $T_2$ distribution predictions using 50 simulated datasets and the same metrics described above for quantitative evaluation. Additionally, we assessed clinical relevance by computing the two-dimensional MWF map from the predicted and reference $p(T_2)$ by summing from $T_2=10$ms to $T_2=40$ms. 
\begin{equation}
    MWF = \sum^{T_2=40ms}_{T_2=10ms}{p(T_2)}
\end{equation}
The accuracy of the different DNN models and a classical DECAES algorithm \citep{decaes} were compared using the MSE metric for evaluation.

\subsection{Robustness against variations in the TE}
\subsubsection{Robustness against known variations in the TE}

In order to evaluate the robustness of various models against changes in both the TEs and SNR, we generated a test set similar to the one described in Section~\ref{sec:fix_te_1d_test}, with the first echo time randomly varied between 5 ms and 15 ms. 
This allowed us to assess the ability of the models to perform effectively under different acquisition sequences.

\subsubsection{Robustness against unseen TEs}
Finally, we evaluated the models' robustness against unknown variations in TE by simulating a training dataset consisting of 20-echo signals with $TE_{min}$ values varying uniformly between 10-11ms, which we referred to as the $TE\in[10,11]ms$ dataset. We also created a separate test set of 20-echo signals, but with $TE_{min}$ values varying uniformly between 12-15ms, referred to as the $TE\in[12,15]ms$ dataset.
We compared the performance of four models: the MIML$_{fix}$ model trained with a fixed $TE_{min}$ of 10ms, the MIML$_{var}$ model, and our two variations of the $P_2T_2$ model. The last three models were trained with the $TE\in[10,11]ms$ dataset.
We tested all four models on the $TE\in[12,15]ms$ dataset and evaluated their robustness by computing the mean-squared error (MSE) and the Wasserstein distance between the predicted $T_2$ distribution and the ground truth.

\subsection{Clinical Data}

We demonstrated the clinical applicability of our approach by calculating $T_2$ distributions from healthy volunteers' brain MRI scans of 4. The MRI data of subjects 1-3 were acquired using a 3T MRI scanner (Siemens Prisma) by the Lab for Advanced MRI from Tel-Aviv University \citep{-Eliezer2015MethodSequences}. The acquisition protocol involved a 2D multi-echo spin-echo sequence with the following parameters: acquisition matrix = 128x104; voxel-size = 1.64x1.64x3mm; number of echoes (N-echoes) = 20; flipping-angle = $180^{\circ}$; $\Delta TE=12ms$.
The 4th multi-echo spin-echo (MESE) scan was obtained from a public source \citep{Whitaker2020MyelinMRI}, and was acquired using an initial 90-degree excitation pulse followed by 32 refocusing (180 degrees) pulses, resulting in 32 echoes with $\Delta TE=10ms$, acquisition matrix = 200x200x9; N-echoes = 32. We used the first 20 echoes only to maintain consistency with the acquisition protocol for subjects 1-3, to enable comparison using the same models.

\section{Results}
\label{sec:results}

\subsection{Models accuracy in unseen SNR levels}
\subsubsection{One-dimensional simulated data}
\begin{figure}[h!]
\centering
\includegraphics[width=.49\textwidth]{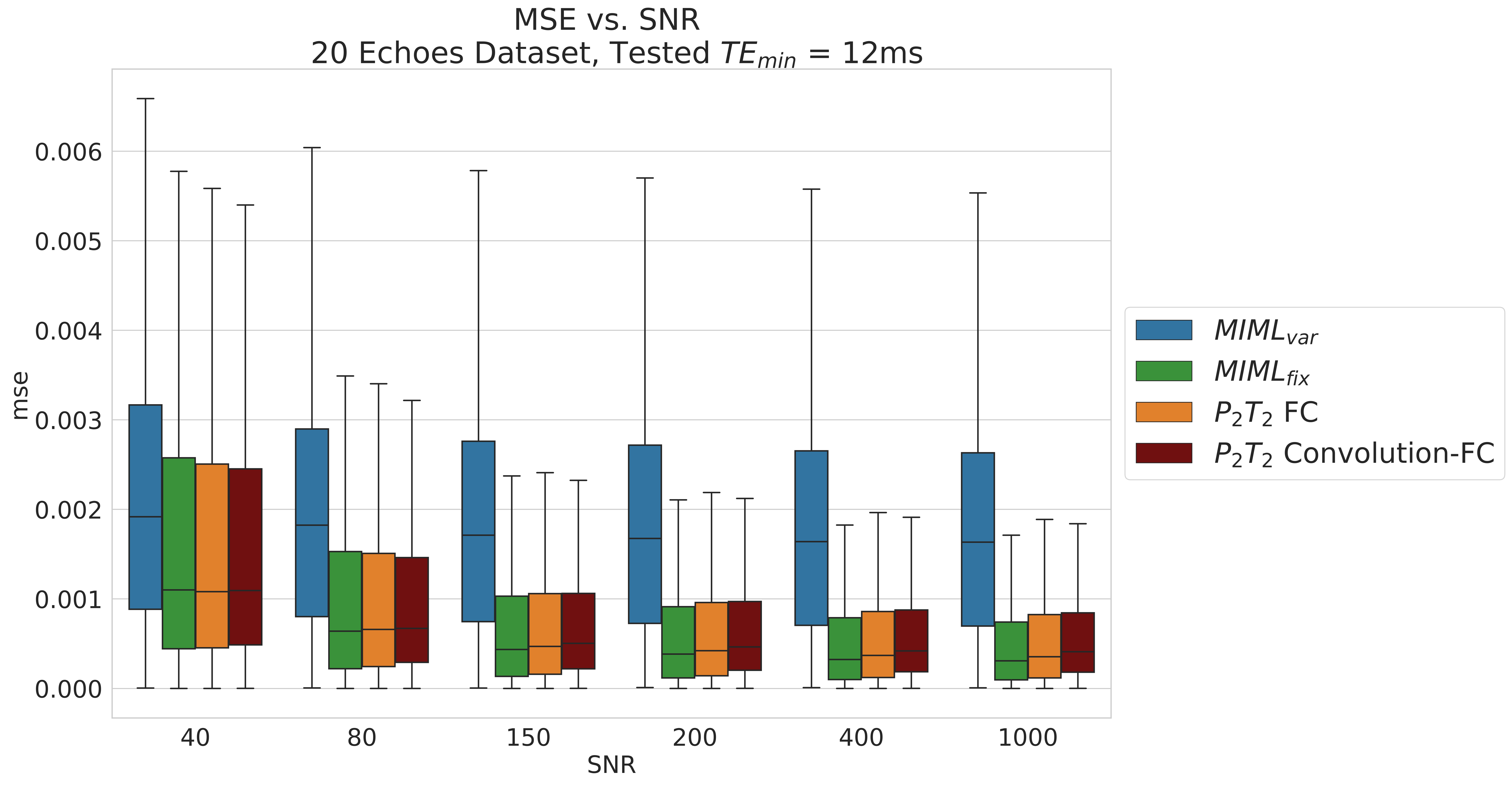}
\hfill
\includegraphics[width=.49\textwidth]{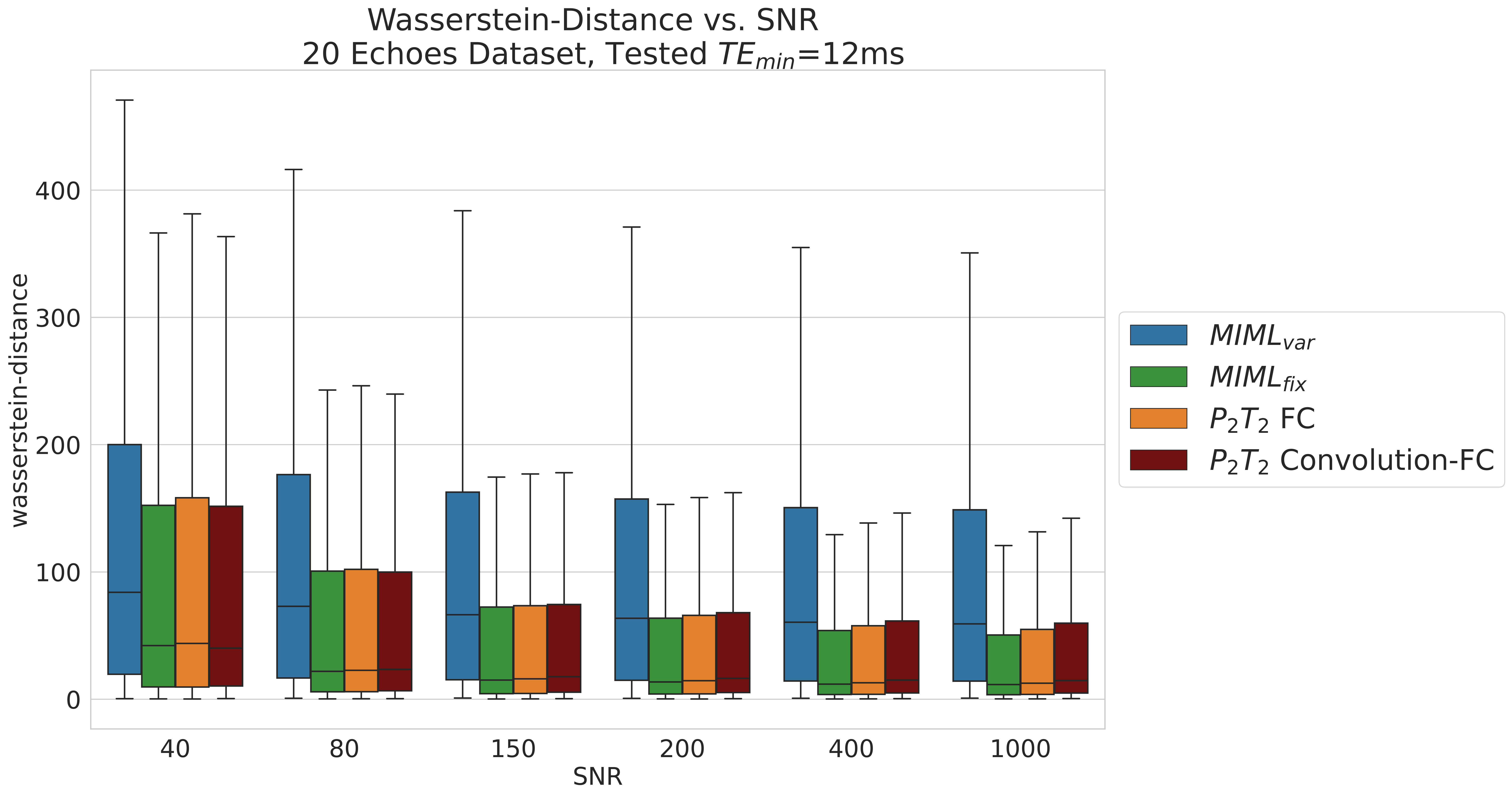}
\caption{\label{fig:boxplot_n20_TEfix} A summary of the models' performances across different signal-to-noise ratios (SNRs) in the context of a fixed-TE test-set ranging from $12$ to $240$ms. The right panel shows the boxplots of the mean squared error (MSE) between the reference and reconstructed distributions, while the left panel displays the Wasserstein distance (WD) between them.}
\end{figure}

Figure~\ref{fig:boxplot_n20_TEfix} presents a summary of the estimation accuracy results for the different models. The boxplots depict the mean squared error (MSE) and Wasserstein distance as a function of signal-to-noise ratio (SNR) during inference. For low SNR levels ($SNR \in [10, 80]$), the $P_2T_2$-FC model achieved lower MSE from the baseline MIML$_{fix}$ model, with a statistical-significant improvement, as shown in Table \ref{tab:te_12_low_snr_ttest}. 

\begin{table}[!]
    \centering
    \begin{tabular}{| *{6}{c|} }
    \hline
    Tested & \multicolumn{2}{c|}{MIML$_{fix}$} & \multicolumn{2}{c|}{$P_2T_2$-FC} & T-test
    \\
     SNR & $\mu(MSE)$ & $\sigma(MSE)$ & $\mu(MSE)$ & $\sigma(MSE)$ & p-value 
     \\
     \hline
    10 & 0.00508 &  0.00436 & 0.00399 & 0.00338 & $p\ll 1e^{-4}$
    \\
    \hline
    20 & 0.00317 &  0.00305 & 0.00291 & 0.00281 & $p\ll 1e^{-4}$ 
    \\
    \hline
    30 & 0.00241 &  0.00255 & 0.00233 & 0.00247 & $p\ll 1e^{-4}$ 
    \\
    \hline
    40 & 0.00198 &  0.00225 & 0.00194 & 0.00219 & $p\ll 1e^{-4}$ 
    \\
    \hline
    80 & 0.00121 &  0.00159 & 0.00120 & 0.00152 & $p\ll 1e^{-4}$ 
    \\
    \hline
    \end{tabular}
    \caption{Summary of Mean-Squared Error (MSE) and One-Sided paired t-test Results. The table provides the average MSE and standard deviation for the predicted $T_2$ distribution, as well as the one-sided t-test p-value indicating the statistical significance of the improvement observed with the $P_2T_2$-FC model enhancement.}
    \label{tab:te_12_low_snr_ttest}
\end{table}

\begin{table}[!]
    \centering
    \begin{tabular}{| *{6}{c|} }
    \hline
    Tested & \multicolumn{2}{c|}{MIML$_{fix}$} & \multicolumn{2}{c|}{$P_2T_2$-FC} & T-test
    \\
     SNR & $\mu(MWF)$ & $\sigma(MWF)$ & $\mu(MWF)$ & $\sigma(MWF)$ & p-value 
     \\
     \hline
    10 & 0.06952 &  0.17453 & 0.03259 & 0.08100 & $p\ll 1e^{-4}$
    \\
    \hline
    20 & 0.01782 &  0.06451 &  0.01044 & 0.02552 & $p\ll 1e^{-4}$ 
    \\
    \hline
    30 & 0.00755 & 0.02484  & 0.00588 & 0.01493 & $p\ll 1e^{-4}$ 
    \\
    \hline
    40 & 0.00472 & 0.0133  & 0.00410 &  0.0107& $p\ll 1e^{-4}$ 
    \\
    \hline
    80 &  0.00190&  0.00524 & 0.00183 &  0.00491& $p\ll 1e^{-4}$ 
    \\
    \hline
    \end{tabular}
    \caption{Summary of Myelin-Water Fraction (MWF) Error and One-Sided paired t-test Results. The table provides the average error and standard deviation for the predicted MWF from the $T_2$ distribution, as well as the one-sided t-test p-value indicating the statistical significance of the improvement observed with the $P_2T_2$-FC model enhancement.}
    \label{tab:te_12_low_snr_ttest}
\end{table}

In higher SNR levels ($SNR \in [80,1000]$), the $P_2T_2$ models exhibited comparable accuracy to the baseline MIML$_{fix}$ model for both the $T_2$ distribution and myelin water fraction (MWF) estimations. Conversely, the MIML$_{var}$ model showed over 50\% higher average error and a wider distribution of errors.

\subsubsection{Two-dimensional brain-based data}

\begin{figure}[h!]
\centering
\includegraphics[width=.47\columnwidth]{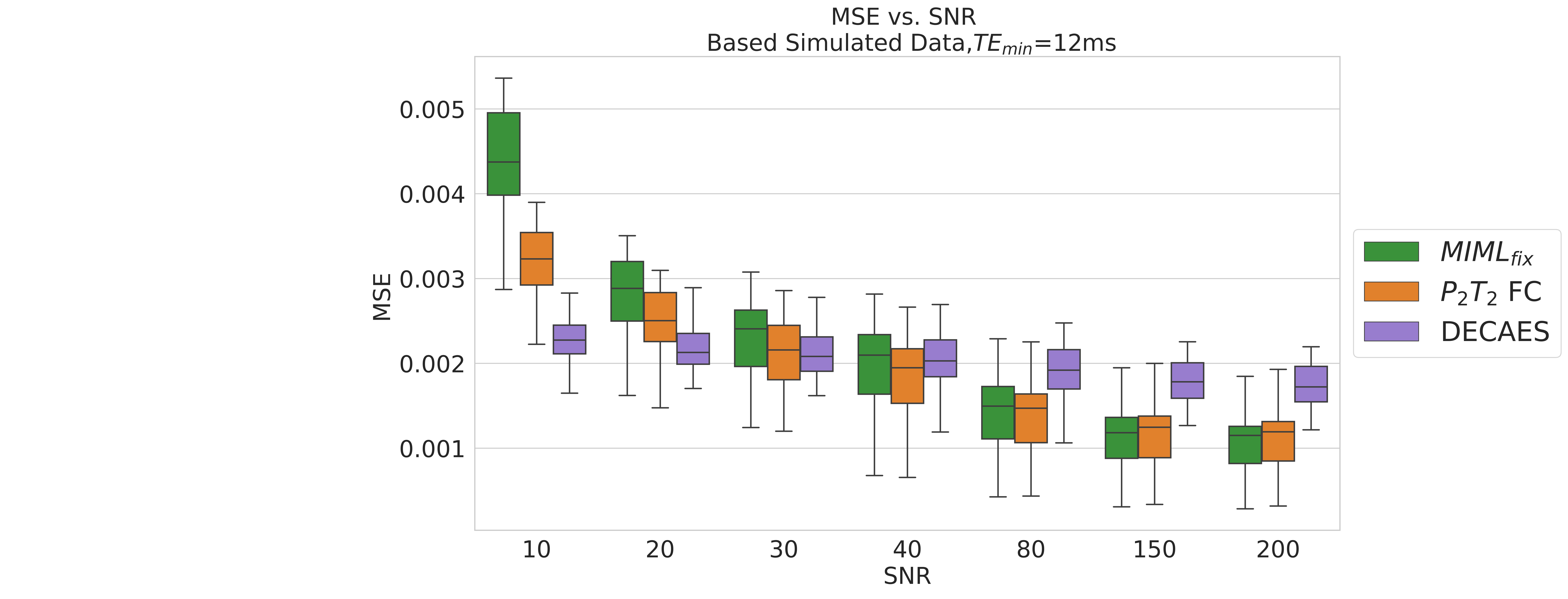}
\includegraphics[width=.47\columnwidth]{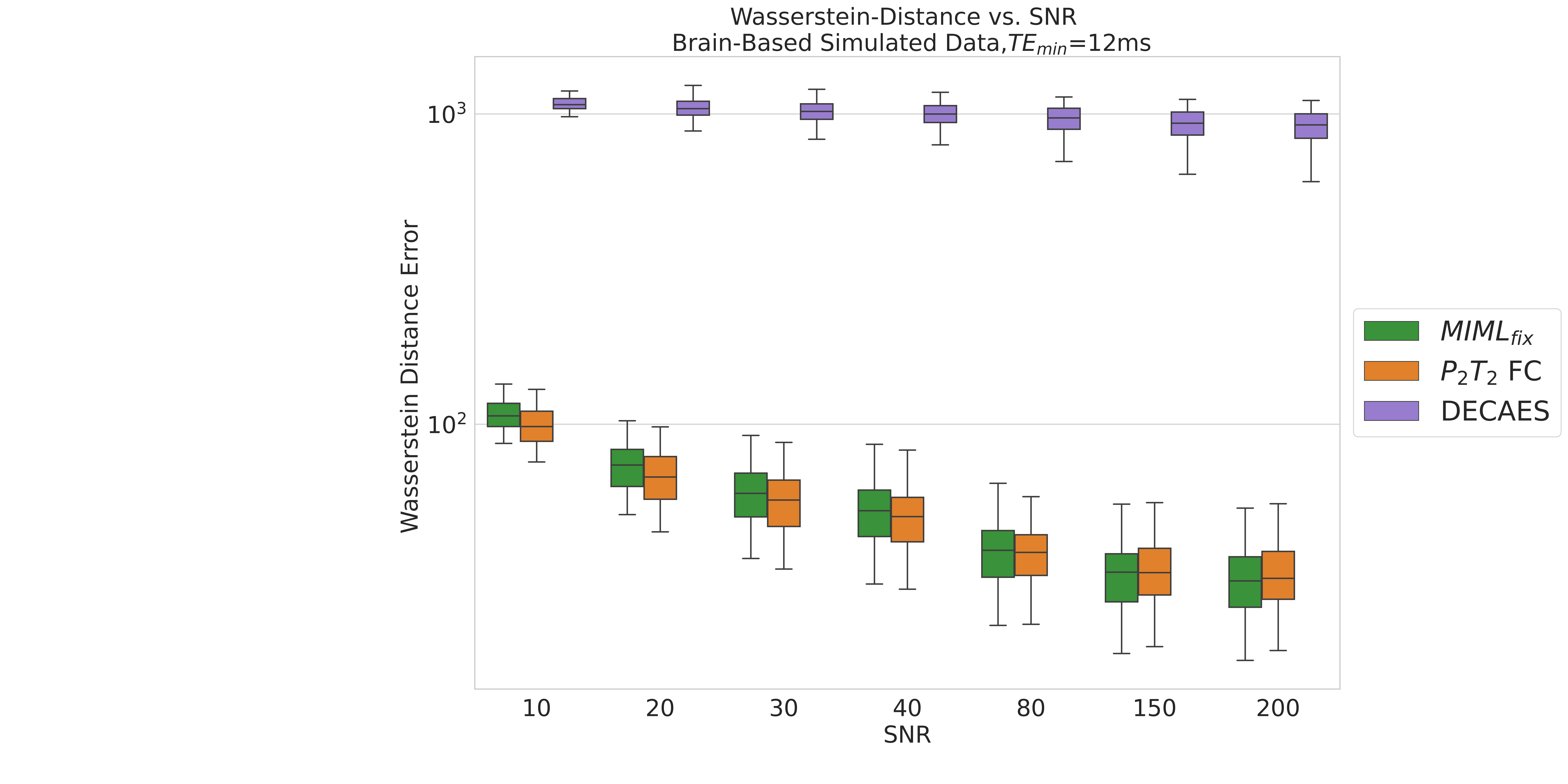}
\caption{\label{fig:oasis_sim_graph}Estimating the $T_2$ distribution accuracy using mean squared error (MSE) and Wasserstein distance, on a simulated brained-based test set. Notably, when testing on low SNR levels that were not seen during training, the $P_2T_2$ model demonstrated superior performance compared to the baseline MIML$_{fix}$ model. 
Although the DECAES model exhibited lower MSE for SNR<40, the Wasserstein distance between its prediction and the actual ground truth suggests inaccurate prediction.}
\end{figure}

Figure~\ref{fig:oasis_sim_graph} presents a comparative analysis of different models' ability to accurately estimate the $T_2$ distribution on a simulated test set based on the OASIS dataset. Our proposed $P_2T_2$ model demonstrated superior performance when tested under low signal-to-noise ratio conditions compared to the baseline MIML$_{fix}$ model.

The DECAES model showed a lower MSE between the predicted $T_2$ distribution and the ground truth, but a relatively high Wasserstein distance, indicating that the prediction accuracy was suboptimal. Upon further analysis of the average signal within several regions of interest, it was observed that the DECAES $p(T_2)$ could not be segregated into two distinct components, which precludes the creation of multi-component relaxation maps (MWF maps).


\begin{figure*}[h!]
\centering
\includegraphics[width=\textwidth]{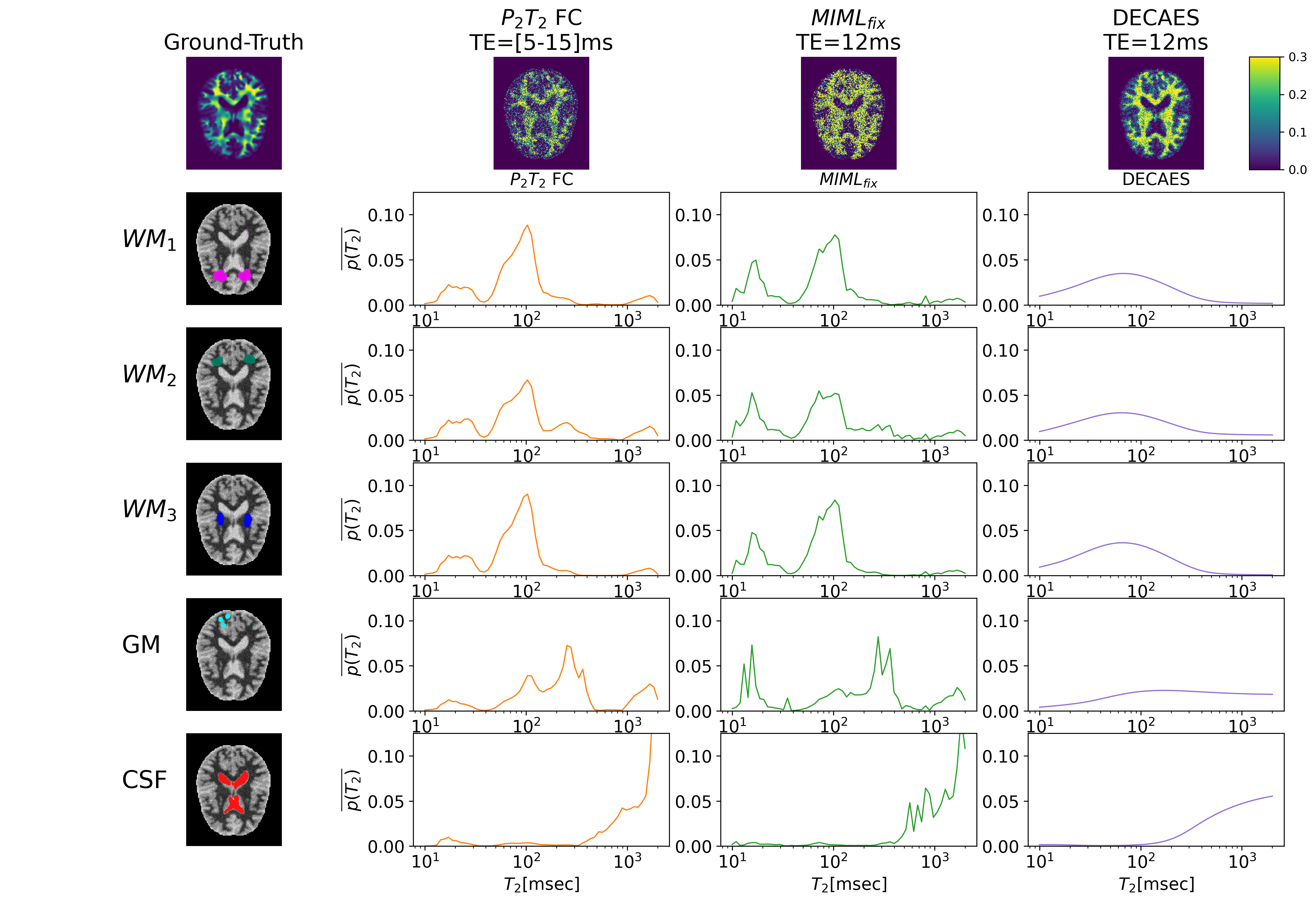}

\caption{\label{fig:OASIS_Simulation_Results_TE=12ms_FA=110.25}Myelin-Water Fraction Prediction for OASIS-Based Simulation, TE=12ms, FA = 110.25. The top row is the MWF prediction from each method. The subsequent rows (2-7) depict the 10th echo from the Multi-Echo Spin Echo (MESE) scan with the corresponding region
of interest (ROI), followed by the average p($T_2$) prediction within the ROI.}
\end{figure*}

Figure~\ref{fig:OASIS_Simulation_Results_TE=12ms_FA=110.25} displays the myelin water fraction (MWF) maps and the average $p(T_2)$ signal within a region of interest (ROI) for a simulated multi-echo MRI image based on the OASIS dataset, with a signal-to-noise ratio (SNR) of 10. The figure illustrates the predictive performance of different models for MWF values, with the $P_2T_2$-FC model outperforming the MIML$_{fix}$ model. Notably, the predicted $T_2$ distributions from the DECAES model lack two components, suggesting a high level of unreliability in the MWF values produced by this approach.

\subsection{Robustness against variations in the TE and the SNR}
\subsubsection{Robustness against known variations in the TE}

Our $P_2T_2$-FC model improved overall $T_2$ reconstruction accuracy by $\sim35\%$, compared to MIML$_{var}$ trained on the same dataset ($18.4e^{-4}$ vs. $27.9e^{-4}$, $p \ll 1e^{-4}$), and the Myelin-Water fraction reconstruction by $28\%$ ($0.00651$ vs. $0.00913$, $p\ll 1e^{-4}$).

\begin{figure}[h!]
\centering
\includegraphics[width=.49\columnwidth]{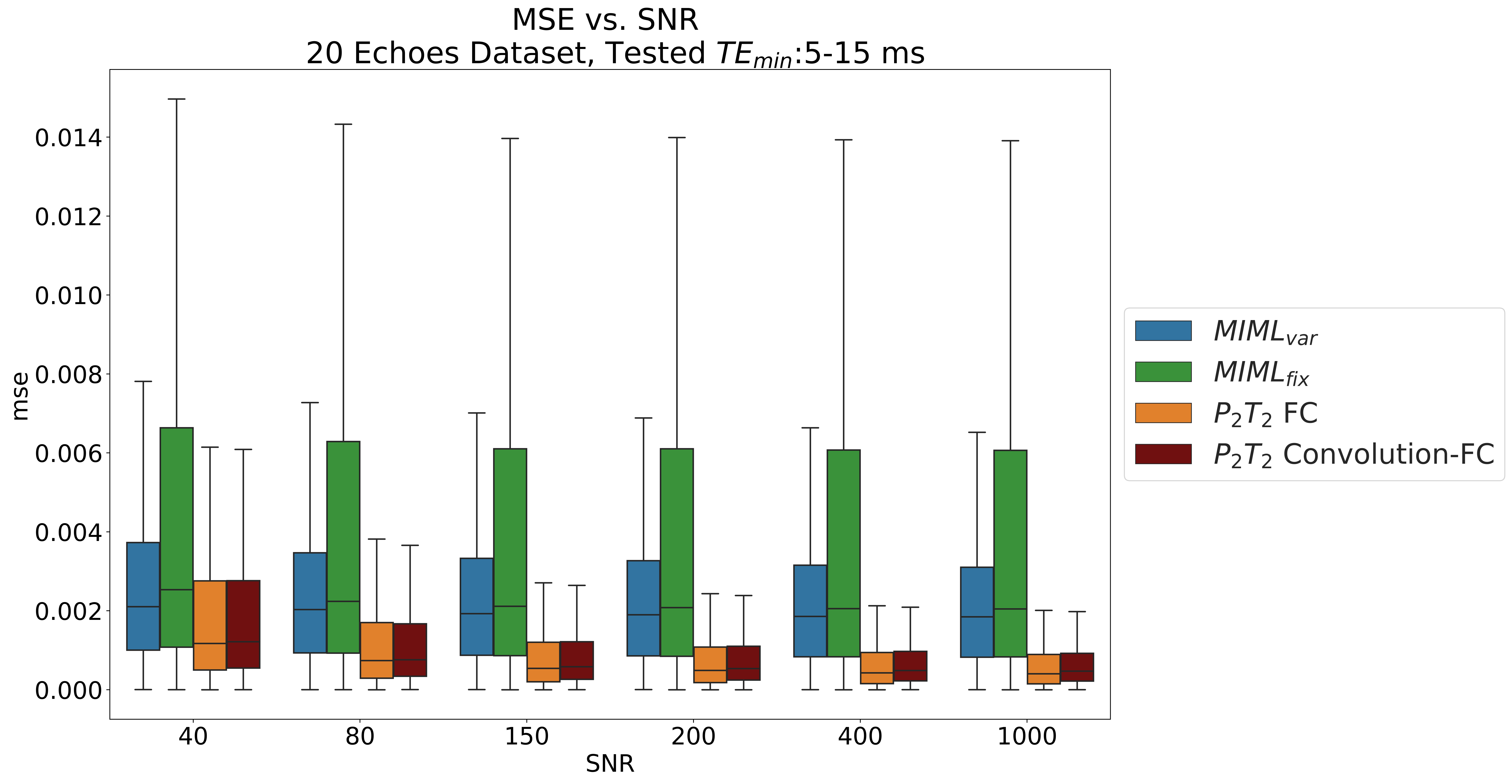}
\hfill
\includegraphics[width=.49\columnwidth]{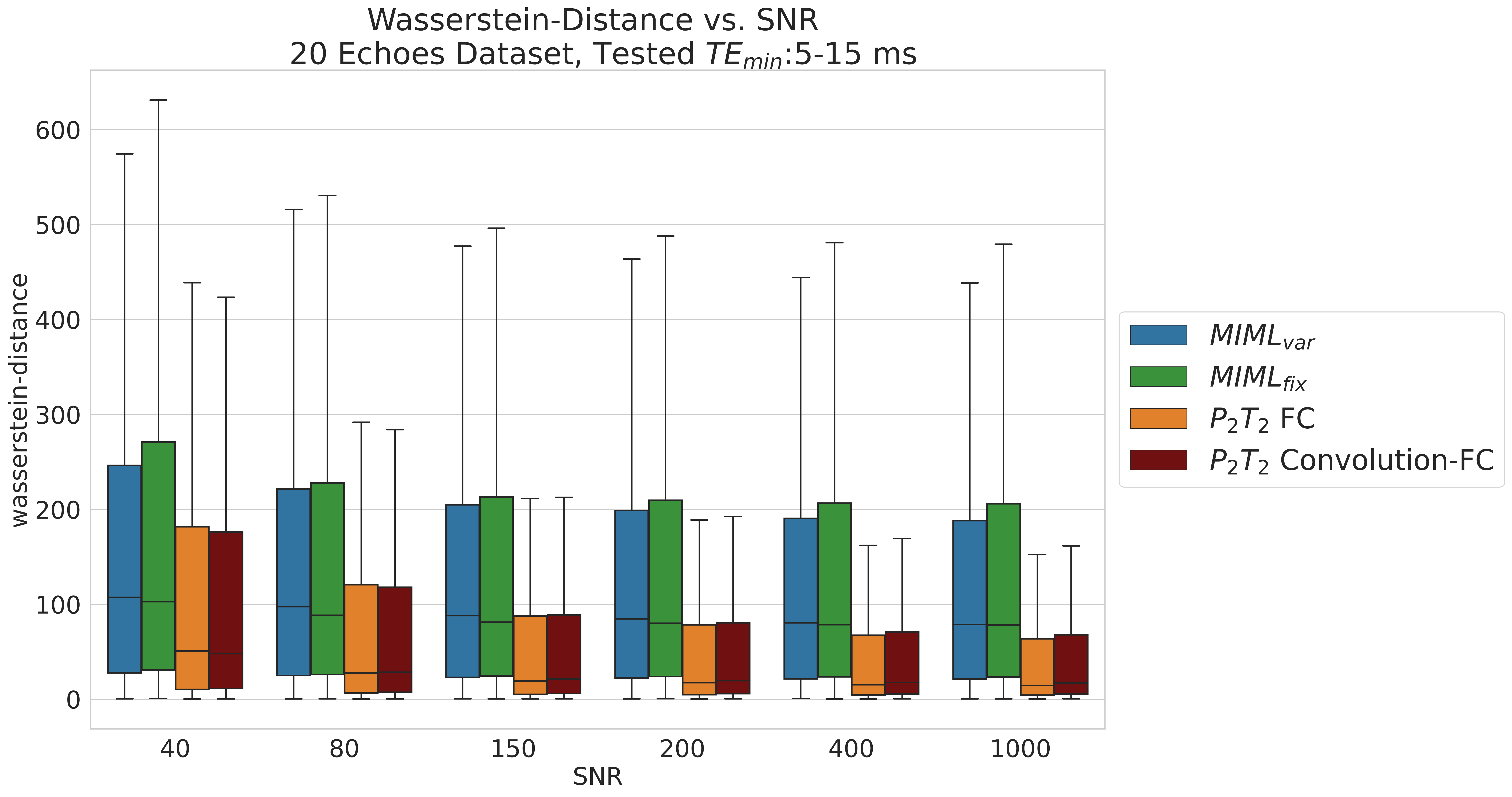}
\caption{\label{fig:boxplot_n20_TEvar}The MSE and Wasserstein distance between the reconstructed and ground truth $T_2$ distributions were compared using boxplots. Right: MSE boxplot comparison for the 4 different models at various tested SNR levels. Left: Wasserstein distance between the predicted $p(T_2)$ and the ground truth for the tested models.}
\end{figure}

Figure \ref{fig:boxplot_n20_TEvar} presents box-plots of the mean-squared error (MSE) and Wasserstein distance, summarizing the robustness of the various models against variations in the acquisition TE. The box-plots depict each model's median and quartile ranges of the error measures. Notably, the $P_2T_2$ models exhibit a smaller average MSE and a narrower range of Q1-Q3 percentiles, suggesting their superior robustness to variations in the acquisition protocols. Conversely, the high error values of the MIML$_{var}$ model imply its instability in the face of varying TEs used during acquisition.

\subsubsection{Robustness against unseen TEs}

\begin{figure}[th!]
\centering
\includegraphics[width=0.49\columnwidth]{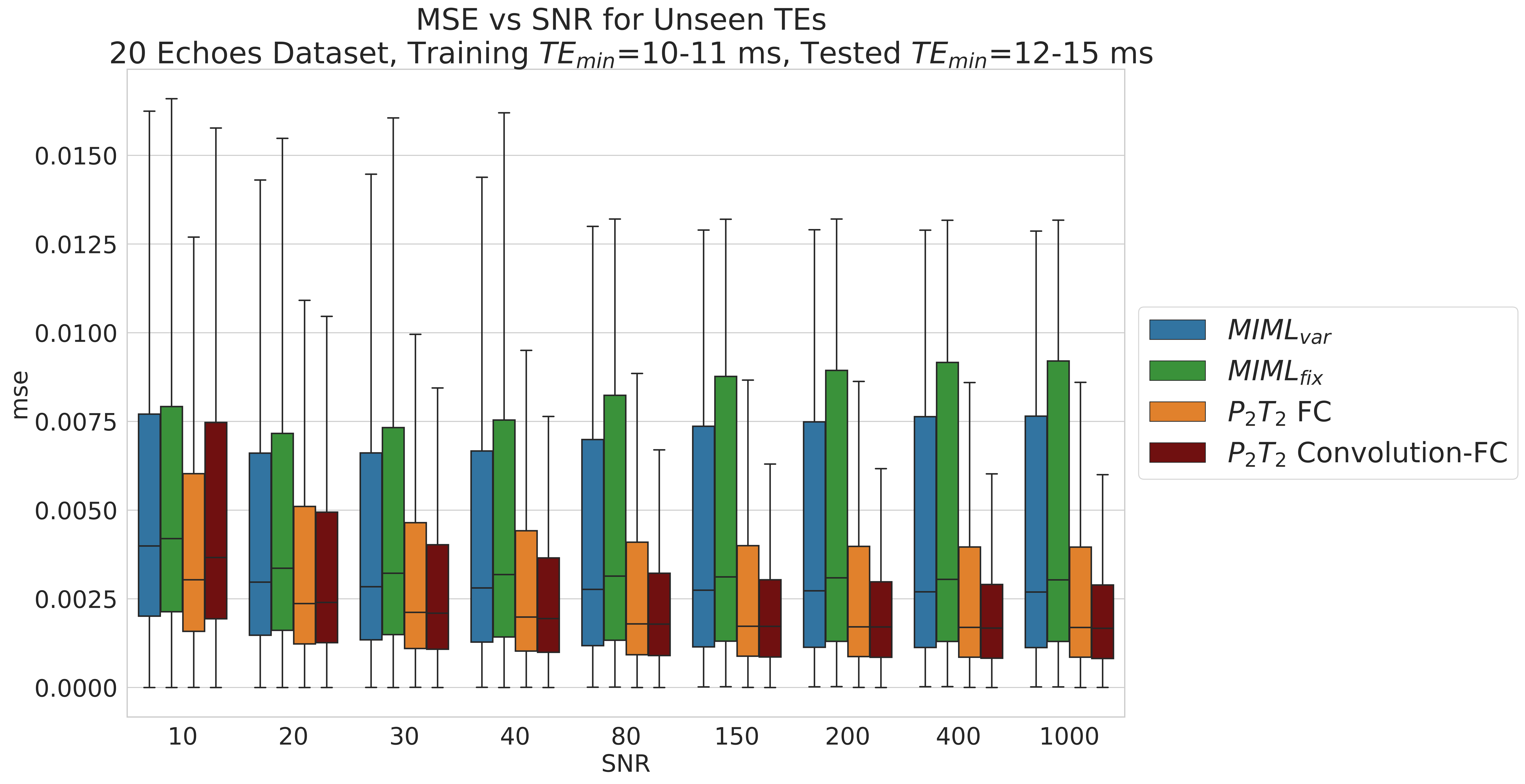}
\hfill     
\includegraphics[width=0.49\columnwidth]{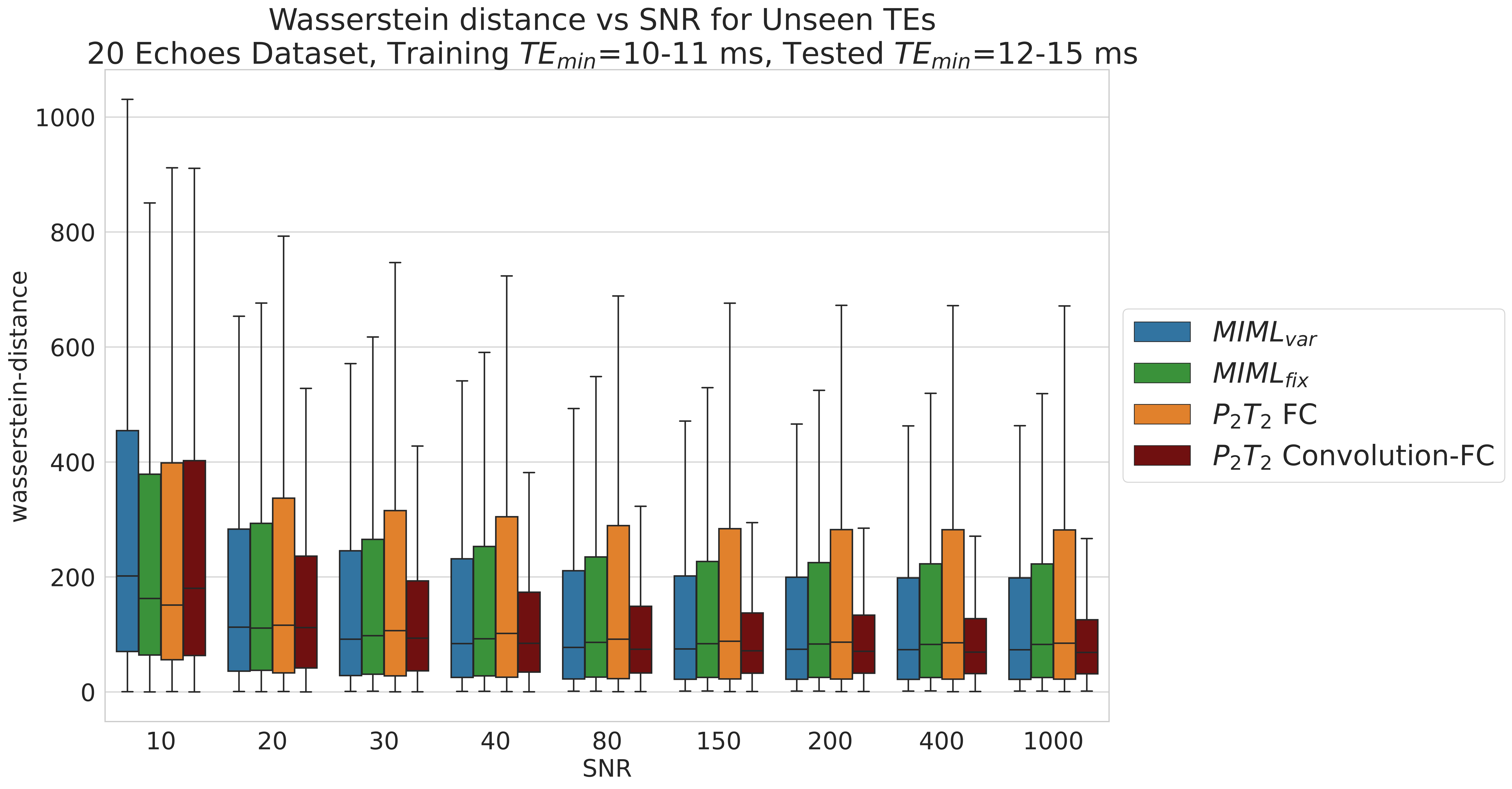}
\caption{\label{unseen_te_boxplots}
Models' performance on TE sequences that were not included in the training. Right: boxplot of the mean-squared error (MSE) between the predicted $p(T_2)$ and the ground truth, as a function of the signal-to-noise ratio (SNR) level. Left: boxplot of the Wasserstein distance between the predicted $p(T_2)$ and the ground truth, as a function of the SNR level.}
\end{figure}

Figure~\ref{unseen_te_boxplots} provides a summary of the models' performance on a test-set comprising previously unseen echo times. The figure highlights that the $P_2T_2$ models exhibited superior performance on new echo times, implying a better aptitude for learning the underlying physical characteristics of the signal-decay model.

\subsection{Real data}
\begin{figure*}[h!]
\centering
\includegraphics[width=\textwidth]{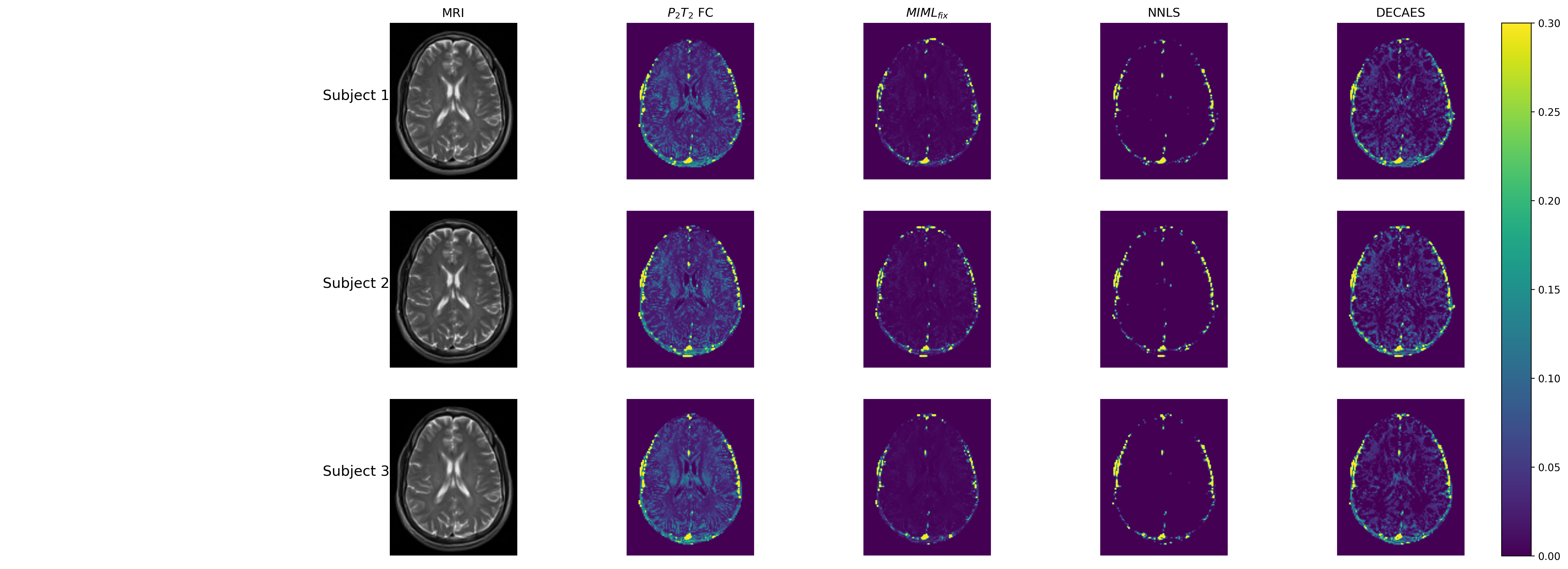}
\caption{\label{fig:subcet1-3}Myelin-Water Fraction (MWF) estimation for multi-echo MRI scans of subjects 1-3, reconstructed using the different models.
The first row depicts a slice from the 20-echoes MRI scan, followed by the MWF reconstruction maps of the $P_2T_2$ model, which achieved the most detailed MWF map.
The MWF maps of the MIML, NNLS, and DECAES models are shown in rows 3-5.}
\end{figure*}

Figure \ref{fig:subcet1-3} displays the $T_2W$ MRI images (at TE=120ms) derived from multi-echo 2D-MRI data with $TE_{min}=12$ ms and a total of 20 echoes for subjects 1-3, along with the corresponding MWF predictions obtained by various models and classical methods. Notably, the MWF map produced by the $P_2T_2$ model shows the most detailed information compared to those generated by the MIML$_{fix}$ model, the DECAES method, and the NNLS algorithm.

Figure \ref{fig:blue_data} illustrates the multi-echo MRI data and MWF predictions for the 4th healthy subject, acquired with a TE of 10ms. In this experiment, we compared the performance of a newly trained MIML$_{fix}$ model, which was trained with $TE=10ms$, with the $P_2T_2$ FC, MIML$_{var}$, and DECAES models.
The $T_2$ distributions for six distinct regions-of-interest (ROI) are presented in rows 2-7. The final column of each row represents the average prediction from each model within the respective ROI. The $P_2T_2$-FC model achieves detailed $T_2$ distributions where the Myelin component can be distinguished from other components. In contrast, the $T_2$ distributions produced by DECAES exhibit a broad lobe with no differentiation between the Myelin component and the IES or the GM components.

\section{Discussion and Conclusion}
\label{sec:discussion}

In this study, we introduced a physically-primed deep neural network (DNN) model for precise estimation of $T_2$ distribution from multi-echo $T_2W$ MRI data, which is robust to diverse acquisition sequences. Our comprehensive evaluation of 1D and 2D synthetic data, as well as 2D and real data, indicated that our physically-informed model outperformed the MIML baseline model trained on data acquired with varying TEs. Additionally, our model achieved more accurate estimations than a baseline MIML model trained on TEs different from those used during inference. Furthermore, our proposed model demonstrated comparable performance to the baseline DNN model trained on a specific TE array used during inference. When applied to real human MRI data, our model generated highly detailed MWF maps with remarkable conformity to anatomical structure compared to other models.

Our findings suggest that incorporating the TEs as a part of the DNN model can enhance its generalization capacity. Consequently, this results in models that are robust to variations in the acquisition process and signal-to-noise (SNR) levels without compromising estimation accuracy compared to models trained using the same acquisition protocol employed during inference. The improved robustness and generalization capacity of our $P_2T_2$ model suggests that it learned the inverse problem related to $T_2$ distribution estimation from multi-echo $T_2W$ MRI data, rather than simply learning a correlation between input and output data.

Our methodology builds upon the latest developments in supervised deep learning techniques for solving inverse problems across various domains. Our central premise is that integrating the system's full-forward model as part of the DNN framework will incentivize the model to generalize the inverse problem instead of merely learning a correlation between input and output data.

Our study highlighted the significant benefits of our physically-primed approach in improving DNN robustness while maintaining state-of-the-art accuracy. However, there are some limitations that need to be acknowledged. Firstly, we only examined two DNN architectures that encode TEs, but other potential architectures, such as a dual encoder model, could be explored. Secondly, our voxel-wise fitting approach to estimate the $T_2$ distribution independently solves the inverse problem for each voxel in the MRI scan, without considering spatial correlations, potentially leading to less smooth maps than methods incorporating CNN models to leverage spatial information \citep{SynthMap}. Lastly, similar to previous DNN-based methods, our model cannot reconstruct a $T_2$ distribution when using sequences with different echoes. For instance, a model trained with 20 echoes could not estimate the $T_2$ distribution of an MRI scan with only 10 echoes.

In conclusion, we have demonstrated the utility of incorporating the system's complete forward model into the DNN architecture, resulting in improved robustness and generalization capabilities against distribution shifts in the acquisition process. Specifically, our approach enables accurate estimation of $T_2$ distributions from multi-echo $T_2W$ MRI data, which could yield valuable imaging biomarkers for diverse clinical applications and facilitate the use of heterogeneous data from different acquisition protocols in large-scale clinical studies.

Furthermore, the benefits of our proposed approach extend beyond this specific application, as it has the potential to enhance DNN robustness in solving inverse problems related to various quantitative MRI challenges and other scientific domains. Our findings have significant implications for advancing the development and implementation of DNN models for real-world applications.

\section*{Acknowledgments}
The study was supported in part by research grants from the United States Israel Bi-national Science Foundation (BSF), the Israel Innovation Authority,  the Israel Ministry of Science and Technology, and  the Microsoft Israel and Israel Inter-University Computation Center program. 

We thank Thomas Yu, Erick Jorge Canales Rodriguez, Marco Pizzolato, Gian Franco Piredda, Tom Hilbert, Elda Fischi-Gomez, Matthias Weigel, Muhamed Barakovic, Meritxell Bach-Cuadra, Cristina Granziera, Tobias Kober, and Jean-Philippe Thiran, from \citep{MIML} for sharing their synthetic data generator with us.
We also thank Prof. Noam Ben-Eliezer and the Lab for Advanced MRI at Tel-Aviv University for sharing the real MRI data with us.

\bibliographystyle{unsrt}

\bibliography{references}

\end{document}